\documentclass[onecolumn,showpacs,preprintnumbers,amsmath,amssymb]{revtex4}
\usepackage{graphicx}
\usepackage{color}
\usepackage{hyperref}
\usepackage{bm}
\usepackage{slashed}

\usepackage{tikz}
\usetikzlibrary{arrows,decorations.pathmorphing,backgrounds,positioning,fit,petri}
\usetikzlibrary{decorations.markings}

\usepackage{CJK}
\begin{document}

\begin{CJK*}{GBK}{fs} 
\CJKtilde

\title{The Calculation of Force in Lattice Quantum Chromodynamics}

\author{Daming Li}
\email{lidaming@sjtu.edu.cn}

\affiliation{School of Mathematical Sciences, Shanghai Jiao Tong
University, Shanghai, 200240, China}

\date{\today}

\begin{abstract}
The calculation of force is most difficult part in lattice Quantum Chromodynamics (QCD). This lecture gives the details of the force calculation in one-loop Symanzik improved action, Wilson fermion with clover term, asqtad fermion, HISQ fermion, rooted staggered fermion, smeared fermion, staggered Wilson fermion, overlap fermion and domain wall fermion. The even-odd precondition are also considered in these calculations.
\end{abstract}



\maketitle

\section{Introduction}

Lattice QCD (LQCD), started in 1974 \cite{Wilson_2445}, is a mature subject,  and it provides a framework in which the strong interactions can be studied from first principles, from low to high energy scales. At the high energy level, it can test the perturbative method as shown in deep inelastic experiments where a very high momentum transfer and  weakly coupled quarks appear as the prominent degrees of freedom. While at the low energy level, only LQCD can provide tool to study non-perturbative phenomena. Now, lattice QCD is an important tool to test the Standard Model, where it can give various hadronic matrix elements, which can compare those obtained using phenomenological approaches. It can also explore the QCD phase diagram for temperature and finite density. There are many other physical applications of LQCD, e.g., for the calculation of hadron spectrum, pseudoscalar decay constants,
kaon bag parameter, semileptonic form factors, strong gauge coupling, quark masses, etc.  Lattice QCD becomes quantitatively predictive only with the advent of supercomputers \cite{Creutz_2308}.
There are several textbooks and notes available for detailed introduction \cite{Montvay}\cite{Gupta_9807028}\cite{Rothe}\cite{Smit}\cite{Gattringer_2009}, where the discretization of the continuum QCD are introduced. The goal of this lecture is to give a concise formula for the force coming from the gauge action and fermion action, which is the most important part in the algorithms of LQCD. Based on these formula, I hope that the reader can understand the free LQCD codes (such as MILC, Chroma, CPS, etc.) easily.

The arrangement of the paper is as follows. In section \ref{one_loop},
the one-loop Symanzik improved action is presented and its force calculation is given in details. In section \ref{Wilson}, Wilson fermion action with clover term and its fermion force is given. The fermion force for asqtad fermion (Sec. \ref{asqtad}), HISQ fermion (Sec. \ref{HISQ}), smeared fermions (Sec. \ref{smeared}), rooted staggered fermion (Sec. \ref{rooted}), staggered Wilson fermion (Sec. \ref{staggered_wilson}), overlap fermion (Sec.\ref{overlap}) and domain wall fermion (Sec.\ref{domain_wall}) are calculated in details, respectively.
Conclusions are given in section \ref{conclusion}.

\section{One-loop Symanzik improved action}\label{one_loop}

\begin{figure}[h]
\centering
\includegraphics[width=8cm,height=3cm]{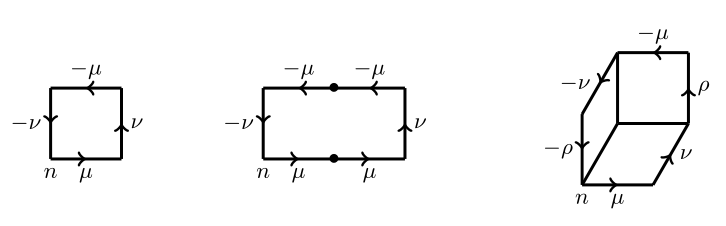}
\caption{the plaquette (Left), rectangle (Middle) and cube loop (Right) are shown.} \label{21_12_4_5}
\end{figure}

The one-loop Symanzik improved action is based on the sum of the plaquette, rectangle and cube loops \cite{Alford_87}:
\begin{eqnarray}\label{18_3_27_30}
\sum_n \sum_{P_n } c_P P_n & = & \sum_n \sum_{1\leq\mu<\nu\leq 4}U_{n,\mu}U_{n+\hat\mu,\nu}U_{n+\hat\mu+\hat\nu,-\mu}U_{n+\hat\nu,-\nu} - \nonumber \\ &&
\frac{1+0.4805 \alpha_s}{20 u_0^2}
\sum_n \sum_{\mu\neq \nu=1,\cdots,4}U_{n,\mu}U_{n+\hat\mu,\mu}U_{n+2\hat\mu,\nu}U_{n+2\hat\mu+\hat\nu,-\mu}U_{n+\hat\mu+\hat\nu,-\mu}U_{n+\hat\nu,-\nu} - \nonumber\\ &&   \frac{2\times 0.03325 \alpha_s}{u_0^2}\sum_n \sum_{1\leq\mu<\nu<\rho\leq 4, \pm\nu,\pm\rho}U_{n,\mu}U_{n+\hat\mu,\nu}U_{n+\hat\mu+\hat\nu,\rho}U_{n+\hat\mu+\hat\nu+\hat\rho,-\mu}U_{n+\hat\nu+\hat\rho,-\nu}U_{n+\hat\rho,-\rho}
\end{eqnarray}
where the sum in each line are shown in FIG. \ref{21_12_4_5}.  

 We always use the lattice unit $a=1$ in the whole paper.
Here $\sum_n$ denotes the sum over all sites $n$ in 4D lattice and $1\leq \mu, \nu, \rho\leq 4$ denote the four directions of the 4D lattice. In the third line of (\ref{18_3_27_30}) there are 16 terms: $\{1\leq\mu<\nu<\rho\leq 4, \pm\nu,\pm\rho\}= \{(1,\pm 2, \pm 3), (1,\pm 2, \pm 4), (1,\pm 3, \pm 4), (2,\pm 3, \pm 4)\}$. Here $+\mu$ and $-\mu$ denotes the positive and negative direction of $\mu$, respectively. For each site $n$, there are 34 paths including 6 plaquette, 12 rectangles and 16 cube loops, which begin and terminate at $n$. Denote by $P_n$ any one in these 34 paths. We also use $P_n$ to represent the multiplication of SU(3) matrix $U$ along this path $P_n$, and denote by $c_P$ the corresponding coefficient which depends on the path type. In particular, $c_P$ does not depends on lattice site $n$. As usual the link variable $U_{n,\mu}$ is the SU(3) matrix defined on the link
$(n,n+\hat\nu)$ and $U_{n+\hat\nu,-\nu} = U_{n,\nu}^\dagger$. In general, $P_n$ with $L=4,6$ links has a form
\begin{eqnarray}\label{18_6_6_3_0}
P_n =  U_{n_1,d_1}\cdots  U_{n_L,d_L}
\end{eqnarray}
where $n_1=n_L+\hat d_L=n$, $n_{i+1}=n_i+\hat d_i$, $i=1,\cdots,L-1$. It is noted that the $L$ directions $\{d_i\}_{i=1}^L$ are determined by the type of path $P$. $u_0$ in (\ref{18_3_27_30}) is the tadpole coefficient and $\alpha_s= -\frac{4\log(u_0)}{3.06839}$.
   The one-loop Symanzik improved action is
\begin{eqnarray}\label{18_6_4_1}
S_G[U] = \frac{\beta}{3} \sum_n \sum_{P_n } c_P \Big[ 3 - \text{Re} \text{tr}(P_n) \Big] 
=   \frac{\beta}{3} \sum_{\text{all 34 paths } P} \sum_n  c_P\Big[ 3 - \text{Re} \text{tr}(P_n) \Big]
\end{eqnarray}
where $\text{Re} \text{tr}$ denotes the real part of the trace of $3\times 3$ matrix, $\beta$ is the inverse gauge coupling. The one-loop Symanzik improved action has the lattice artifacts of $O(\alpha_s a^2)$ which does not include the one-loop contributions from dynamical fermions.
The standard Wilson gauge action is replaced by this perturbatively-improved action with tadpole-improved correction terms that remove the leading errors due to the lattice.
Let $\{T_i\}_{i=1}^8$ be the 8 generators of Lie algebra $su(3)$ of SU(3). These generators are traceless, complex, and hermitian $3 \times 3$ matrices obeying the normalization condition
\begin{eqnarray}\label{18_1_29_3_0}
\text{tr}[T_jT_k] =\frac{1}{2}\delta_{jk},\quad j,k=1,\cdots,8
\end{eqnarray}
Each link variable can be represented by
\begin{eqnarray}\label{18_1_29_3}
U_{n,\mu} = \exp\Big(\text{i}\sum_{i=1}^8\omega^{i}_{n,\mu} T_i\Big), \quad \mu= 1,2,3,4
\end{eqnarray}
with 8 real numbers $\{\omega^{i}_{n,\mu}\}_{i=1}^8$ for each link $(n,\mu)$. By using the presentation of each link in (\ref{18_1_29_3}), the gauge action in (\ref{18_6_4_1}) is a function of $\omega^{i}_{n,\mu}$. The gauge force of $S_G$ is
\begin{eqnarray}\label{18_7_25_1}
-\sum_i T_i \frac{\partial S_G}{\partial \omega_{k,\rho}^i}
& = &  \frac{\beta}{3} \sum_{\text{all 34 paths } P}  c_P\sum_n \sum_i T_i \ \text{Re} \text{tr}\Big(\frac{\partial P_n}{\partial \omega^i_{k,\rho}} \Big) \nonumber \\
& = &  \frac{\beta}{3} \sum_{\text{all 34 paths } P}  c_P\sum_{l=1}^L\sum_n \sum_i T_i \ \text{Re} \text{tr}\Big(U_{n_1,d_1} \cdots U_{n_{l-1},d_{l-1}} \frac{\partial U_{n_l,d_l}}{\partial \omega^i_{k,\rho}}  U_{n_{l+1},d_{l+1}} \cdots U_{n_L,d_L} \Big)
\end{eqnarray}
where $P_n$ is given by (\ref{18_6_6_3_0}).
Since
\begin{eqnarray}\label{18_7_26_1}
\frac{\partial U_{n_l,d_l}}{\partial \omega^i_{k,\rho}} =   \left\{ \begin{array}{ll}
\text{i} T_i U_{n_l,d_l}, &   \text{if } U_{n_l,d_l} = U_{k,\rho}, \text{  i.e. },  n_l=k \text{ and } d_l=\rho  \\
 U_{n_l,d_l}(-\text{i} T_i), & \text{if } U_{n_l,d_l} = U_{k,\rho}^\dagger \equiv U_{k+\hat\rho,-\rho}, \text{  i.e.  },  n_l=k+\hat\rho \text{ and } d_l=-\rho\\
0, &  \text{otherwise }  \\
\end{array} \right.
\end{eqnarray}
only two terms in the sum over $n$ in (\ref{18_7_25_1}) does not vanish, i.e.,
\begin{eqnarray}\label{18_7_26_2}
n=n_1=k-\sum_{j=1}^{l-1}\hat d_j, \quad \text {if}  \quad (n_l,d_l)=(k,\rho)
\end{eqnarray}
\begin{eqnarray}\label{18_7_26_3}
 n=n_1=k+\hat\rho-\sum_{j=1}^{l-1}\hat d_j, \quad \text {if} \quad (n_l,d_l)=(k+\hat\rho,-\rho)
 \end{eqnarray}
  Inserting (\ref{18_7_26_1}) into (\ref{18_7_25_1}), and using the cyclic property of trace,
  $\text{Re}\text{tr}(B) =  \text{Re}\text{tr} (B^\dagger)$ for any complex matrix $B$, the gauge force in (\ref{18_7_25_1}) then becomes
\begin{eqnarray}\label{18_7_25_2}
&&-\sum_i T_i \frac{\partial S_G}{\partial \omega_{k,\rho}^i} \nonumber \\
& = &  \frac{\beta}{3} \sum_{\text{all 34 paths } P}  c_P\sum_{l=1}^L\Big\{ \sum_i T_i \
\text{Re}   \text{tr} \Big[  (\text{i}T_i) ( U_{n_l,d_l}  \cdots U_{n_L,d_L} U_{n_1,d_1} \cdots U_{n_{l-1},d_{l-1}}) \Big]_{(n_l,d_l)=(k,\rho)} + \nonumber \\
 &&\sum_i T_i \ \text{Re}   \text{tr} \Big[  (\text{i}T_i)  (U_{n_{l},d_{l}}^\dagger U_{n_{l-1},d_{l-1}}^\dagger \cdots U_{n_{1},d_{1}}^\dagger  U_{n_{L},d_{L}}^\dagger \cdots  U_{n_{l+1},d_{l+1}}^\dagger) \Big]_{(n_l,d_l)=(k+\hat\rho,-\rho)}\Big\} \nonumber \\
& = &  \frac{\beta}{6} \sum_{\text{all 34 paths } P}  c_P\sum_{l=1}^L \Big\{
     \text{i} ( U_{n_l,d_l}  \cdots U_{n_L,d_L} U_{n_1,d_1} \cdots U_{n_{l-1},d_{l-1}})_{TA}\Big|_{(n_l,d_l)=(k,\rho)} +\nonumber \\
&&     \text{i} (U_{n_{l},d_{l}}^\dagger U_{n_{l-1},d_{l-1}}^\dagger \cdots U_{n_{1},d_{1}}^\dagger  U_{n_{L},d_{L}}^\dagger \cdots  U_{n_{l+1},d_{l+1}}^\dagger)_{TA}\Big|_{(n_l,d_l)=(k+\hat\rho,-\rho)} \Big\} \nonumber \\
& = &  \frac{\beta}{6} \sum_{\text{all 34 paths } P}  c_P\sum_{l=1}^L \Big\{   \text{i} U_{k,\rho} \Big[
( U_{n_{l+1},d_{l+1}}  \cdots U_{n_L,d_L} U_{n_1,d_1} \cdots U_{n_{l-1},d_{l-1}})\Big|_{(n_l,d_l)=(k,\rho)} +\nonumber \\
&&  ( U_{n_{l-1},d_{l-1}}^\dagger \cdots U_{n_{1},d_{1}}^\dagger  U_{n_{L},d_{L}}^\dagger \cdots  U_{n_{l+1},d_{l+1}}^\dagger)\Big|_{(n_l,d_l)=(k+\hat\rho,-\rho)} \Big]  \Big\}_{TA}
\end{eqnarray}
where $B_{TA}$ is the traceless and anti-Hermitian part of $3\times 3$ complex matrix $B$
\begin{eqnarray}\label{18_6_6_11}
B_{TA} = \frac{B-B^\dagger}{2} - \frac{\text{tr}(B-B^\dagger)}{6}\text{I}_3
\end{eqnarray}
In (\ref{18_7_25_2}), we used
\begin{eqnarray}\label{18_6_6_13}
  \sum_{i=1}^8 T_i  \text{Retr}[\text{i} T_i B ]=  \frac{1}{2}\sum_i T_i \Big\{ \text{tr} [\text{i} T_i B]  + \text{c.c.} \Big\}  =
\frac{1}{2} \sum_i T_i  \text{tr}\Big[\text{i} T_i ( B - B^\dagger )  \Big]  =  \sum_i T_i   \text{tr}\Big[\text{i} T_i  B_{TA} \Big] =  
\frac{1}{2}  \text{i} B_{TA}
\end{eqnarray}
for any $3\times 3$ complex matrix $B$. We always use $\text{c.c.}$ to represent the complex conjugate.  

\begin{figure}[h]
\centering
\includegraphics[width=6cm,height=5cm]{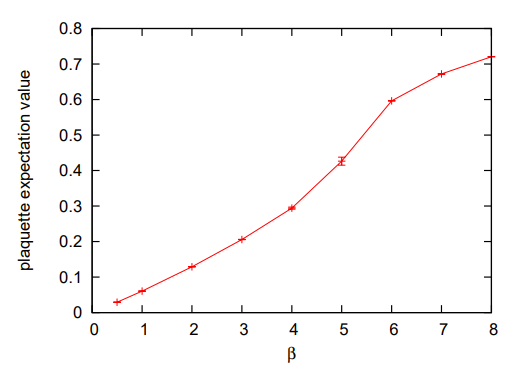}
\includegraphics[width=7cm,height=5cm]{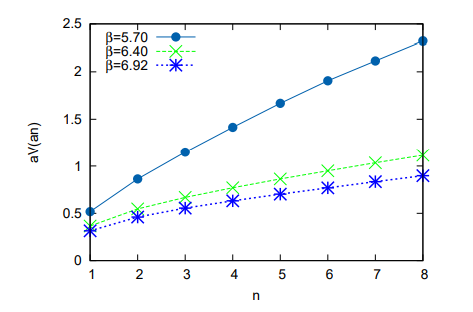}
\caption{The plaquette expectation and static quark potential $aV(an)$ depends on the inverse gauge coupling $\beta$.} \label{21_12_9_1}
\end{figure}

In summary, the gauge fermion force has two contributions. One contribution comes from the paths passing through the link $(k,\rho)$ in the positive direction and the other from the paths passing through the link $(k,\rho)$ in the negative direction.
The implementation of gauge force calculation can be written as follows.
\begin{enumerate}
\setlength{\parskip}{-3pt}
\item For all direction $\rho$
\item \hspace{0.5cm} $t_{k,\rho} =0$ for each site $k$
\item \hspace{0.5cm} For all 34 global path $P$
\item \hspace{1.2cm} For each link $l$ of $P$
\item \hspace{1.9cm} If $d_l\neq \pm \rho$, go to step 4
\item \hspace{1.9cm} If $d_l=\rho$, $ t_{k,\rho} \ +=c_PU_{n_{l+1},d_{l+1}}\cdots U_{n_{L},d_{L}}U_{n_1,d_1} \cdots U_{n_{l-1},d_{l-1}}$ for all sites   $k=n_l$.
\item \hspace{1.9cm} If $d_l=-\rho$, $ t_{k,\rho} \ +=c_PU_{n_{l-1},d_{l-1}}^\dagger \cdots U_{n_{1},d_{1}}^\dagger  U_{n_{L},d_{L}}^\dagger \cdots  U_{n_{l+1},d_{l+1}}^\dagger$ for all sites $k=n_l-\hat \rho$.
\item \hspace{0.5cm} Calculate $\frac{\beta}{6} \text{i}(U_{k,\rho} t_{k,\rho})_{TA}$ for each site $k$
\end{enumerate}
Based on the standard Wilson gauge action, we use the hybrid monte carlo algorithm to simulate the plaquette expectation value and 
static quark potential $aV(an)$, as shown in FIG. \ref{21_12_9_1} where the lattice size in the space directions and Euclidean time direction are 16 and 6, respectively.

\section{Wilson fermion action with clover term}\label{Wilson}
The Wilson fermion matrix with clover term is \cite{Sheikholeslami_216}
\begin{eqnarray}\label{18_7_17_2}
D= A - \kappa {\cal D}
\end{eqnarray}
 with the diagonal part
\begin{eqnarray}\label{18_7_17_3}
A_{n,m} = \Big[1 - \frac{c_{\text{sw}}}{4( m+ 4)}\sum_{1\leq\mu<\nu\leq 4} [\gamma_\mu,\gamma_\nu]  F_{n;\mu\nu}\Big]\delta_{n,m}
\end{eqnarray}
and non-diagonal part in lattice space
\begin{eqnarray}\label{18_7_17_4}
{\cal D}_{n,m} = \sum_{\mu= 1}^{4}\Big( (1-\gamma_\mu)  U_{n,\mu}\delta_{n+\hat\mu,m}+
 (1+\gamma_\mu)  U_{n,-\mu}\delta_{n-\hat\mu,m}\Big)
\end{eqnarray}
Here $\{\gamma_\mu\}_{\mu=1}^4$ are the $4\times 4$ Gamma matrices, $m$ is the non-dimensional bare mass of each specie, $\kappa =\frac{1}{2(m+4)}$, $c_{sw}$ is the  Sheikholeslami-Wohlert coefficient, which can be determined by perturbative method \cite{Luscher_219}, tadpole improvement \cite{Lepage_219}\cite{DeGrand_219} and nonperturbative method \cite{Jansen_221}.  The clover part $F_{n;\mu\nu}$ in (\ref{18_7_17_3}) is anti-Hermitian
\begin{eqnarray}\label{17_7_20_0_0}
 F_{n;\mu\nu} = \frac{1}{8}(Q_{n;\mu\nu}-Q_{n;\nu\mu})
\end{eqnarray}
with
\begin{eqnarray}\label{17_7_20_8}
Q_{n;\mu\nu} = U_{n; \mu,\nu}+ U_{n;\nu,-\mu} + U_{n;-\mu,-\nu} +U_{n;-\nu,\mu} = Q_{n;\nu\mu}^\dagger
\end{eqnarray}
where $U_{n; \mu,\nu}= U_{n,\mu}U_{n+\hat\mu,\nu}U_{n+\hat\mu+\hat\nu,-\mu}U_{n+\hat\nu,-\nu} = U_{n; \nu,\mu}^\dagger$, etc. Thus the diagonal part in (\ref{18_7_17_3})
is
\begin{eqnarray}\label{18_7_18_0}
A_{n,m} = \Big[1 - \frac{c_{\text{sw}}}{32( m+ 4)}\sum_{1\leq \mu\neq\nu\leq 4} [\gamma_\mu,\gamma_\nu]  Q_{n;\mu\nu}\Big]\delta_{n,m}
\end{eqnarray}

 The partition function of LQCD for the Wilson fermion with two degenerate flavors reads
\begin{eqnarray}\label{18_1_29_1}
Z = \int {\cal D} [U] e^{-S_G[U]} (\det D)^2   = \int {\cal D} [U] e^{-S_G[U]}  \det (D^\dagger D)
= \int {\cal D} [U] {\cal D} [\phi] e^{-S[U]}
\end{eqnarray}
where we used $\det D = \det D^\dagger$ and introduced the pseduofermion (real field) $\phi$. The effective action in (\ref{18_1_29_1}) is
\begin{eqnarray}\label{18_1_29_2}
 S [U]= S_G[U] + \phi^\dagger (D^\dagger D)^{-1}\phi
\end{eqnarray}
The one-loop Symanzik improved action $S_G$ in \ref{18_1_29_2} is given in (\ref{18_6_4_1}) and it's derivative has been calculated in section \ref{one_loop}. The derivative of the fermion action is calculated as follows
\begin{eqnarray}\label{18_6_24_0}
- \frac{\partial}{\partial \omega^{i}_{k,\rho}} [\phi^\dagger (D^\dagger D)^{-1}\phi ]
 =   Y^\dagger  \frac{\partial D}{\partial \omega^{i}_{k,\rho}}  X + \text{c.c.} =
   Y_n^\dagger  \frac{\partial D_{n,m}}{\partial \omega^{i}_{k,\rho}}  X_m + \text{c.c.}
\end{eqnarray}
with
\begin{eqnarray}\label{18_6_24_0_1}
X =  (D^\dagger D)^{-1}\phi, \quad Y = D X
\end{eqnarray}
Since
\begin{eqnarray}\label{18_1_29_19}
\frac{\partial {\cal D}_{n,m}}{\partial \omega^{i}_{k,\rho}} =  (1-\gamma_\rho) (\text{i} T_i) U_{k,\rho}\delta_{n+\hat\rho,m}\delta_{n,k}+
 (1+\gamma_\rho)  U_{k,\rho}^\dagger (-\text{i}T_i) \delta_{n-\hat\rho,m}\delta_{m,k}
\end{eqnarray}
the contribution of (\ref{18_1_29_19}) to (\ref{18_6_24_0}) is
\begin{eqnarray}\label{18_1_29_20_1}
 &&  Y_n^\dagger  \frac{\partial  {\cal D}_{n,m}}{\partial \omega^{i}_{k,\rho}}  X_m + \text{c.c.} \nonumber \\
 &=& \Big[  Y_k^\dagger  (1-\gamma_\rho) (\text{i} T_i) U_{k,\rho} X_{k+\hat\rho}
- Y_{k+\hat\rho}^\dagger (1+\gamma_\rho)  U_{k,\rho}^\dagger (\text{i}T_i)  X_k\Big] + \text{c.c.}\nonumber \\
 &=&   \text{tr} \Big(\text{i}T_i (B - C)\Big)+ \text{c.c.}
\end{eqnarray}
with the $3\times 3$ matrix
\begin{eqnarray}\label{18_1_29_20_2}
 B=  \Big((1-\gamma_\rho)U_{k,\rho} X_{k+\hat\rho}\Big)\otimes Y_k^\dagger, \quad \text{i.e., }
    B_{b,a} = \Big((1-\gamma_\rho)U_{k,\rho} X_{k+\hat\rho}\Big)_{\alpha b} (Y_k^\dagger)_{\alpha a}
\end{eqnarray}
\begin{eqnarray}\label{18_1_29_20_3}
 C = X_{k}\otimes  \Big(Y_{k+\hat\rho}^\dagger (1+\gamma_\rho)  U_{k,\rho}^\dagger\Big)
\end{eqnarray}
The last equality in (\ref{18_1_29_20_1}) use the fact 
\begin{eqnarray}\label{18_1_29_20_4}
Y_k^\dagger  (1-\gamma_\rho) (\text{i} T_i) U_{k,\rho} X_{k+\hat\rho}
 = Y_k^\dagger (\text{i} T_i) Z  = Y^\dagger_{k,\alpha a} (\text{i} T_i)_{ab} Z_{\alpha b}   
 =\text{tr}(\text{i}T_i B)
\end{eqnarray}
with $Z =  (1-\gamma_\rho) U_{k,\rho} X_{k+\hat\rho}$. 
We always use $a,b,c,d, e,f$ etc. for the color indices and $\alpha,\beta$ etc. for the Dirac indices. The contribution of ${\cal D}$ for the
fermion force $-\sum_i T_i \frac{\partial}{\partial \omega^{i}_{\rho}(k)} [\phi^\dagger (D^\dagger D)^{-1}\phi ]
$ is
\begin{eqnarray}\label{18_7_17_30}
-\sum_i T_i \Big( \text{tr} (\text{i}T_i (B - C))+ \text{c.c.}\Big)  = -\text{i}(B-C)_{TA}
\end{eqnarray}

By using (\ref{18_7_18_0}), the contribution of $A$ to (\ref{18_6_24_0}) is
\begin{eqnarray}\label{18_7_17_31}
 &&  Y_n^\dagger  \frac{\partial   A_{n,m}}{\partial \omega^{i}_{k,\rho}}  X_m + \text{c.c.} \nonumber \\
 &=&  Y_n^\dagger  \frac{\partial  }{\partial \omega^{i}_{k,\rho}} \sum_{\mu\neq\nu}[\gamma_\mu,\gamma_\nu]  Q_{n;\mu\nu}  X_n + \text{c.c.} \nonumber \\
 &=& \sum_{n=k,k+\hat\rho,k\pm\hat\nu,k+\hat\rho\pm\hat\nu} Y_n^\dagger  [\gamma_\rho,\gamma_\nu] \frac{\partial}{\partial \omega^{i}_{k,\rho}}( Q_{n;\rho\nu} - Q_{n;\nu\rho}) X_n + \text{c.c.}
\end{eqnarray}
where we omit the writing of factor $- \frac{c_{\text{sw}}}{32( m+ 4)}$ before the sum. In the last equality we used
\begin{eqnarray}\label{18_7_17_31_1}
 \frac{\partial  }{\partial \omega^{i}_{k,\rho}} \sum_{\mu\neq\nu}[\gamma_\mu,\gamma_\nu]  Q_{n;\mu\nu}
=\sum_\nu [\gamma_\rho,\gamma_\nu] \frac{\partial}{\partial \omega^{i}_{k,\rho}}( Q_{n;\rho\nu} - Q_{n;\nu\rho})
\end{eqnarray}
where the sum over $n$ and $\nu$ is understood. The sum over $n$ in (\ref{18_7_17_31}) runs for $n=k,k+\hat\rho,k\pm\hat\nu,k+\hat\rho\pm\hat\nu$, otherwise the derivative $\frac{\partial}{\partial \omega^{i}_{k,\rho}}( Q_{n;\rho\nu} - Q_{n;\nu\rho})$ in (\ref{18_7_17_31}) vanishes.  Let us consider the term $n=k+\hat\nu$ and the other terms are similar. Since
\begin{eqnarray}\label{18_7_18_1_1}
&& \frac{\partial}{\partial \omega^{i}_{k,\rho}}( Q_{k+\hat\nu;\rho\nu} - Q_{k+\hat\nu;\nu\rho})  \nonumber \\
&=& \frac{\partial}{\partial \omega^{i}_{k,\rho}}\Big( U_{k+\hat\nu,-\nu}U_{k,\rho}U_{k+\hat\rho,\nu}U_{k+\hat\rho+\hat\nu,-\rho} - U_{k+\hat\nu,\rho}U_{k+\hat\rho+\hat\nu,-\nu}U_{k+\hat\rho,-\rho}U_{k,\nu}\Big)  \nonumber \\
&=&  U_{k+\hat\nu,-\nu}(\text{i}T_i)U_{k,\rho}U_{k+\hat\rho,\nu}U_{k+\hat\rho+\hat\nu,-\rho} - U_{k+\hat\nu,\rho}U_{k+\hat\rho+\hat\nu,-\nu}U_{k+\hat\rho,-\rho}(-\text{i}T_i)U_{k,\nu}
\end{eqnarray}
the contribution to (\ref{18_7_17_31}) is
\begin{eqnarray}\label{18_7_18_1}
 Y_{k+\hat\nu}^\dagger  [\gamma_\rho,\gamma_\nu] \frac{\partial}{\partial \omega^{i}_{k,\rho}}( Q_{k+\hat\nu;\rho\nu} - Q_{k+\hat\nu;\nu\rho}) X_{k+\hat\nu} + \text{c.c.}= \text{tr} \Big( \text{i}T_i (B+C)\Big) + \text{c.c.}
\end{eqnarray}
where
\begin{eqnarray}\label{18_7_26_9}
 B = \Big(U_{k,\rho}U_{k+\hat\rho,\nu}U_{k+\hat\rho+\hat\nu,-\rho}X_{k+\hat\nu} \Big) \otimes \Big(Y_{k+\hat\nu}^\dagger [\gamma_\rho,\gamma_\nu]U_{k+\hat\nu,-\nu} \Big)
\end{eqnarray}
\begin{eqnarray}\label{18_7_26_10}
 C = \Big(U_{k,\nu}X_{k+\hat\nu} \Big) \otimes \Big(Y_{k+\hat\nu}^\dagger [\gamma_\rho,\gamma_\nu]U_{k+\hat\nu,\rho}U_{k+\hat\rho+\hat\nu,-\nu}U_{k+\hat\rho,-\rho} \Big)
\end{eqnarray}
The above calculation shows that $B$ comes from the part of the fermion action
$$Y_{k+\hat\nu}^\dagger [\gamma_\rho,\gamma_\nu]U_{k+\hat\nu,-\nu} U_{k,\rho}U_{k+\hat\rho,\nu}U_{k+\hat\rho+\hat\nu,-\rho}X_{k+\hat\nu}$$
and it's derivative with respect to $\omega^{i}_{k,\rho}$ will insert $\text{i}T_i$ before $U_{k,\rho}$ and thus separate two vectors in $B$ as shown in (\ref{18_7_26_9}). The formula of $C$ in (\ref{18_7_26_10}) can also be understand.

The elements of non-diagonal part in Wilson fermion matrix are nonzero only for two neighbouring sites, and thus the even-odd precondition can be used, i.e., we first label all even sites and then the odd sites. Using these labelling of sites, the Wilson fermion matrix in (\ref{18_7_17_2}) can be written as
\begin{eqnarray*}
D =   \left( \begin{array}{cc}
A_e &  -\kappa D_{eo}\\
-\kappa D_{oe}  & A_o  \\
\end{array} \right)
\end{eqnarray*}
where $D_{eo} = -D_{oe}^\dagger $ and the matrix $D_{eo}$ has the form of (\ref{18_7_17_4}), which is defined on the even sites (rows) and odd sites (columns). $A_e(A_o)$ are the diagonal part $A$, which defined on the even (odd) sites.
 Introducing the Schur complement of $D$
 \begin{eqnarray}\label{18_7_26_15_1}
M = A_e - \kappa^2 D_{eo}A_o^{-1}D_{oe}
\end{eqnarray}
 one has $\det D =\det M \det A_o$. The partition function in (\ref{18_1_29_1}) can be written as
\begin{eqnarray}\label{18_7_16_10}
Z = \int {\cal D} [U] e^{-S_G[U]}  \det (M^\dagger M) (\det(A_o))^2
= \int {\cal D} [U] {\cal D} [\phi] e^{-S[U]}
\end{eqnarray}
where we used $A_o^\dagger = A_o$, the pseudofermion $\phi$ is defined on the even sites and the effective action reads
\begin{eqnarray}\label{18_7_16_11}
 S [U]= S_G[U] + \phi^\dagger (M^\dagger M)^{-1}\phi - 2\text{tr} \ln A_o
\end{eqnarray}
The derivative of the fermion action is
\begin{eqnarray}\label{18_7_16_11_1}
   \frac{\partial}{\partial \omega^{i}} \Big[\phi^\dagger (M^\dagger M)^{-1}\phi \Big]  = Y_e^\dagger  \frac{\partial M}{\partial \omega^{i}}  X_e + \text{c.c.}
\end{eqnarray}
where
\begin{eqnarray}\label{18_7_16_11_2}
 X_e =  (M^\dagger M)^{-1}\phi, \quad Y_e = M X_e
\end{eqnarray}
are   defined on the even sites. From the definition of $M$ in (\ref{18_7_26_15_1})
\begin{eqnarray}\label{18_7_16_11_3}
  Y_e^\dagger  \frac{\partial M}{\partial \omega^{i}}  X_e   =Y_e^\dagger \frac{\partial A_e}{\partial \omega^{i}} X_e  - \kappa^2 \Big( Y_e^\dagger  \frac{\partial D_{eo}}{\partial \omega^{i}}X_o - Y_o^\dagger \frac{\partial A_o}{\partial \omega^{i}}X_o + Y_o^\dagger \frac{\partial D_{oe}}{\partial \omega^{i}}   X_e\Big)
\end{eqnarray}
with $X_o =  A_o^{-1}D_{oe} X_e, Y_o = A_o^{-1}D_{oe} Y_e$ defined on the odd sites. The calculation of
$Y_e^\dagger \frac{\partial A_e}{\partial \omega^{i}} X_e$ ($Y_o^\dagger \frac{\partial A_o}{\partial \omega^{i}}X_o$) is similar to
(\ref{18_7_17_31}) except these formula are define on the even (odd) sites. The calculation of $Y_e^\dagger  \frac{\partial D_{eo}}{\partial \omega^{i}}X_o$ and $ Y_o^\dagger \frac{\partial D_{oe}}{\partial \omega^{i}} X_e$ are also similar to (\ref{18_1_29_20_1}).

The derivative of the action $\text{tr} \ln A_o$ can be calculated as follows:
\begin{eqnarray}\label{18_7_18_10}
 &&  \frac{\partial  \text{tr} \ln A_o}{\partial \omega^{i}_{k,\rho}}  =  \sum_{\text{odd sites } n}\text{tr} \Big( A_{o,n}^{-1} \frac{\partial   A_{o,n}}{\partial \omega^{i}_{k,\rho}} \Big) \nonumber \\
 &=&  \sum_{\text{odd sites } n}\text{tr} \Big( A_{o,n}^{-1} \frac{\partial  }{\partial \omega^{i}_{k,\rho}} \sum_{\mu\neq\nu} [\gamma_\mu,\gamma_\nu]  Q_{n;\mu\nu}\Big) \nonumber \\
 &=& \sum_{\text{odd sites } n=k,k+\hat\rho,k\pm\hat\nu,k+\hat\rho\pm\hat\nu} \text{tr} \Big[ A_{o,n}^{-1}  [\gamma_\rho,\gamma_\nu] \frac{\partial}{\partial \omega^{i}_{k,\rho}}( Q_{n;\rho\nu} - Q_{n;\nu\rho}) \Big]
\end{eqnarray}
which is rather similar to (\ref{18_7_17_31}). For example, if $k$ is the even site, then there are the contribution for the odd site $n=k+\hat\nu$
\begin{eqnarray}\label{18_7_18_11}
\text{tr} \Big[ A_{o,k+\hat\nu}^{-1}  [\gamma_\rho,\gamma_\nu] \frac{\partial}{\partial \omega^{i}_{k,\rho}}( Q_{k+\hat\nu;\rho\nu} - Q_{k+\hat\nu;\nu\rho}) \Big] = \text{tr} \Big( \text{i}T_i (B+C)\Big)
\end{eqnarray}
where
$$ B = \Big(U_{k,\rho}U_{k+\hat\rho,\nu}U_{k+\hat\rho+\hat\nu,-\rho}\Big) \Big(A_{o,k+\hat\nu}^{-1}  [\gamma_\rho,\gamma_\nu]U_{k+\hat\nu,-\nu}\Big) $$
$$ C = \Big(U_{k,\nu}\Big)\Big( A_{o,k+\hat\nu}^{-1}  [\gamma_\rho,\gamma_\nu]U_{k+\hat\nu,\rho}U_{k+\hat\rho+\hat\nu,-\nu}U_{k+\hat\rho,-\rho}\Big) $$

\section{Asqtad fermion}\label{asqtad}

\begin{figure}[h]
\centering
\includegraphics[width=12cm,height=3cm]{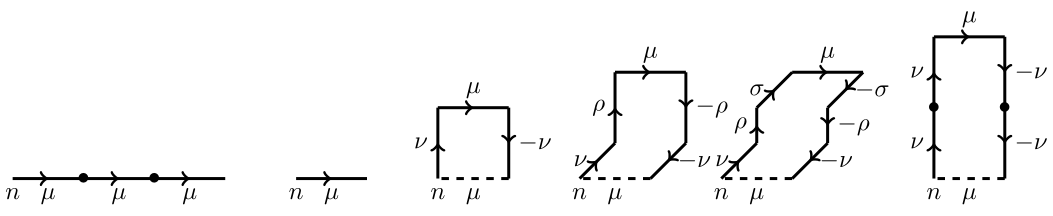}
\caption{The Naik term, one-link, three-staple, five-staple, seven-staple and Lepage term (from left to right).} \label{21_12_4_7}
\end{figure}

Compared to Wilson fermion the staggered fermions are numerically very fast to simulate. This is because the staggered fermion, with only
one component per lattice site, and the massless limit protected by a remnant chiral symmetry. But the major drawbacks is the taste violations due to the exchange of  ultraviolet gluons between different taste components living on neighboring lattice sites.  
The asqtad (a-squared tadpole improved) fermion is introduced largely to reduce this taste violation \cite{Lepage_074502}, which was used in many large scale simulations \cite{Bazavov_1349}. As shown in
in FIG. \ref{21_12_4_7}, 
the asqtad fermion matrix is 
\begin{eqnarray}\label{18_7_18_12}
D= m + c_ND_N + c_1D_1 + c_3D_3 +c_5D_5 +c_7D_7 +c_LD_L
\end{eqnarray}
where $c_N$ etc. are the coefficients, the Gamma matrices in the Wilson fermion are replaced by the staggered factor $\eta_{n,\mu} = (-1)^{n_1+\cdots,n_{\mu-1}}$ for lattice site $n=(n_1,\cdots,n_4)$, $\eta_{n,1} =1$. The asqtad fermion matrix $D$ incldue the ($4\times 2 =8$) Naik terms \cite{Naik_238}
\begin{eqnarray}\label{18_6_6_3}
[D_N]_{n,m} &=& \sum_\mu \eta_{n,\mu} \Big[U_{n,\mu}U_{n+\hat\mu,\mu}U_{n+2\hat\mu,\mu}\delta_{n+3\hat\mu,m} -( \mu \rightarrow -\mu)  \Big]
\end{eqnarray}
the ($4\times 2 =8$) one-link terms
\begin{eqnarray}\label{18_6_6_3_1}
[D_1]_{n,m} &=& \sum_\mu \eta_{n,\mu} \Big[U_{n,\mu}\delta_{n+\hat\mu,m}
 -( \mu \rightarrow -\mu)  \Big]
\end{eqnarray}
the ($4\times 3 \times 4 =48$) three-staple terms
\begin{eqnarray}\label{18_6_6_4}
[D_3]_{n,m} &=& \sum_\mu \eta_{n,\mu} \sum_{\nu\neq \mu}\Big[U_{n,\pm\nu}U_{n\pm\hat\nu,\mu}U_{n+\hat\mu\pm\hat\nu,\mp\nu}\delta_{n+\hat\mu,m}
 -( \mu \rightarrow -\mu)  \Big]
\end{eqnarray}
the ($4\times 3 \times 2 \times 8 =192$) five-staple terms
\begin{eqnarray}\label{18_6_6_5}
[D_5]_{n,m} &=& \sum_\mu \eta_{n,\mu} \sum_{\nu\neq \mu}\sum_{\rho\neq \mu,\nu}\Big[U_{n,\pm\nu}U_{n\pm\hat\nu,\pm\rho}U_{n\pm\hat\nu\pm\hat\rho,\mu}U_{n+\hat\mu\pm\hat\nu\pm\hat\rho,\mp\rho}
U_{n+\hat\mu\pm\hat\nu,\mp\nu}\delta_{n+\hat\mu,m}
 -( \mu \rightarrow -\mu)  \Big]
\end{eqnarray}
the ($4\times 3 \times 2 \times 16 =384$) seven-staple terms
\begin{eqnarray}\label{18_6_6_6}
[D_7]_{n,m} &=& \sum_\mu \eta_{n,\mu} \sum_{\nu\neq \mu}\sum_{\rho\neq \mu,\nu}\sum_{\sigma\neq \mu,\nu,\rho}\Big[U_{n,\pm\nu}U_{n\pm\hat\nu,\pm\rho}U_{n\pm\hat\nu\pm\hat\rho,\pm\sigma}U_{n\pm\hat\nu\pm\hat\rho\pm\hat\sigma,\mu}
\nonumber \\ &&
U_{n+\hat\mu\pm\hat\nu\pm\hat\rho\pm\hat\sigma,\mp\sigma}
U_{n+\hat\mu\pm\hat\nu\pm\hat\rho,\mp\rho}
U_{n+\hat\mu\pm\hat\nu,\mp\nu}\delta_{n+\hat\mu,m}  -( \mu \rightarrow -\mu)  \Big]
\end{eqnarray}
and the ($4\times 3 \times 4 =48$) Lepage terms \cite{Lepage_074502}
\begin{eqnarray}\label{18_6_6_7}
[D_L]_{n,m} &=& \sum_\mu \eta_{n,\mu} \sum_{\nu\neq \mu}\Big[U_{n,\pm\nu}U_{n\pm\hat\nu,\pm\nu}U_{n\pm2\hat\nu,\mu}
U_{n+\hat\mu\pm 2\hat\nu,\mp\nu}
U_{n+\hat\mu\pm\hat\nu,\mp\nu}
\delta_{n+\hat\mu,m}  -( \mu \rightarrow -\mu)  \Big]
\end{eqnarray}

 The asqtad fermion matrix (\ref{18_7_18_12}) include the contribution from two neighbouring
sites
\begin{eqnarray}\label{18_6_6_7_1}
V_{n,\mu}\delta_{n+\hat\mu,m} + V_{n,-\mu}\delta_{n-\hat\mu,m}
\end{eqnarray}
where $V= {\cal F}U$ is the fat link depending on the thin link $U$
\begin{eqnarray}\label{18_6_6_7_2}
V_{n,\mu}&=& \eta_{n,\mu} \Big\{ c_1 U_{n,\mu}  +  \nonumber  \\   && c_3 \sum_{\nu\neq \mu}U_{n,\pm\nu}U_{n\pm\hat\nu,\mu}U_{n+\hat\mu\pm\hat\nu,\mp\nu}
+  \nonumber  \\ && c_5 \sum_{\nu\neq \mu}\sum_{\rho\neq \mu,\nu}U_{n,\pm\nu}U_{n\pm\hat\nu,\pm\rho}U_{n\pm\hat\nu\pm\hat\rho,\mu}U_{n+\hat\mu\pm\hat\nu\pm\hat\rho,\mp\rho}
U_{n+\hat\mu\pm\hat\nu,\mp\nu} +  \nonumber  \\  &&c_7 U_{n,\pm\nu}U_{n\pm\hat\nu,\pm\rho}U_{n\pm\hat\nu\pm\hat\rho,\pm\sigma}U_{n\pm\hat\nu\pm\hat\rho\pm\hat\sigma,\mu}
U_{n+\hat\mu\pm\hat\nu\pm\hat\rho\pm\hat\sigma,\mp\sigma}
U_{n+\hat\mu\pm\hat\nu\pm\hat\rho,\mp\rho}
U_{n+\hat\mu\pm\hat\nu,\mp\nu}\Big\}
\end{eqnarray}

The partition function for two degenerate staggered fermions is
\begin{eqnarray}\label{18_7_18_20_1}
Z = \int {\cal D} [U] e^{-S_G[U]}\det (D^\dagger D)
= \int {\cal D} [U] {\cal D} [\phi] e^{-S[U]}
\end{eqnarray}
with the effective action
\begin{eqnarray}\label{18_7_18_21}
 S [U]= S_G[U] + \phi^\dagger (D^\dagger D)^{-1}\phi
\end{eqnarray}
The derivative of the fermion action $\phi^\dagger (D^\dagger D)^{-1}\phi$ is given in (\ref{18_6_24_0}) and (\ref{18_6_24_0_1}). We gather all paths ($8+8+48+192+384+48=688$) in asqtad matrix. All paths can start any site $n$ and
end some site $m = n\pm \hat\mu, n\pm 3\hat\mu$. Let $P$ be these paths with length $L=1,3,5,7$ and denote also by $P$ the multiplication of link variables along $P$: $P = U_1\cdots U_l\cdots U_L$. It's contrbution to
\begin{eqnarray}\label{18_7_18_21_1}
   Y_n^\dagger  \frac{\partial  D_{n,m}}{\partial \omega^{i}_{k,\rho}}  X_m + \text{c.c.}
\end{eqnarray}
 in (\ref{18_6_24_0}) is
\begin{eqnarray}\label{18_7_19_0}
\eta_{n,\mu} \sum_{l=1}^L   Y_n^\dagger U_1\cdots U_{l-1}\frac{\partial  U_l}{\partial \omega^{i}_{k,\rho}} U_{l+1}\cdots U_L X_{m} + \text{c.c.}=   \sum_{l=1}^L \Big[ \text{tr} (\text{i}T_i B )+ \text{c.c.}\Big]
\end{eqnarray}
with
\begin{eqnarray}\label{18_7_19_1}
 B= \text{sign } \times \left\{
  \begin{array}{l }
(U_l U_{l+1}\cdots U_L X_{m}) \otimes (Y_n^\dagger U_1\cdots U_{l-1}), \quad \text{if } U_l = U_{k,\rho}\\
( U_{l+1}\cdots U_L X_{m}) \otimes (Y_n^\dagger U_1\cdots U_{l-1}U_l), \quad \text{if } U_l = U_{k,\rho}^\dagger\\
   \end{array} \right.
\end{eqnarray}
Here $\text{sign} = (\pm)\eta_{n,\mu}$ since this path starts from site $n$ and ends at $n\pm\hat\mu, n\pm3\hat\mu$.
The implementation of staggered fermion force is similar to the algorithm in section \ref{one_loop}.

Similar to the even-odd precondition for the Wilson fermion, the asqtad fermion matrix in (\ref{18_7_18_12}) under the even-odd precondition is written as
\begin{eqnarray*}
D =   \left( \begin{array}{cc}
m &  D_{eo}\\
D_{oe}  & m  \\
\end{array} \right)
\end{eqnarray*}
where
\begin{eqnarray}\label{18_7_19_3_1}
D_{eo} = c_ND_N+\sum_{i=1,3,5,7}c_iD_i+c_LD_L
\end{eqnarray}
  is defined on the even sites (rows) and odd sites (columns). Then $\det D = \det M$ with
$M = m^2 - D_{eo}D_{oe}$.  The partition function is
\begin{eqnarray}\label{18_7_19_2}
Z = \int {\cal D} [U] {\cal D} [\phi] e^{-S[U]}
\end{eqnarray}
with the effective action
\begin{eqnarray}\label{18_7_19_3}
 S [U]= S_G[U] + \phi^\dagger (M^\dagger M)^{-1}\phi
\end{eqnarray}
where $\phi$ is defined on the even sites. The calculation of the fermion force is similar to (\ref{18_7_16_11_1})(\ref{18_7_16_11_2})(\ref{18_7_16_11_3}) where $A_e$, $A_o$ and $\kappa$ are replaced by $m$ and $1$, respectively. $Y_e^\dagger  \frac{\partial D_{eo}}{\partial \omega^{i}}X_o$ is obtained from (\ref{18_7_19_0}) and (\ref{18_7_19_1}) where $n$ is the even site. Similarly, we can calculate
$Y^\dagger_o  \frac{\partial D_{oe}}{\partial \omega^{i}}X_e$.

\section{HISQ fermion}\label{HISQ}
The taste violations in the asqtad action can be further reduced by additional smearings. This is the highly improved staggered quark (HISQ) fermion, which was introduced in \cite{Follana_054502}. At tree-level it removes both $O(a^2)$ errors and, to lowest order in the
quark speed $v/c$, $O(a^4m^4)$ errors. It also substantially reduces effects caused by taste-symmetry
breaking. This makes it attractive not only for light quarks, but means that it is also quite accurate
for heavy quarks. It is being used to directly simulate charm
quarks and to approach direct simulations of bottom quarks \cite{Davies_114504}\cite{McNeile_031503}\cite{Donald_094501}.
  
To introduce the HISQ fermion, we reunitarize the link variable $V$ in (\ref{18_6_6_7_2}) by
\begin{eqnarray}\label{18_7_24_0}
W_{n,\mu} = {\cal U} V_{n,\mu} \equiv  V_{n,\mu} \Big(V_{n,\mu}V_{n,\mu}^{\dagger}\Big)^{-1/2}
\end{eqnarray}
and smear the links $W$ again to obtain $X = {\cal F}W$.
The HISQ fermion matrix  is similar to asqtad fermion matrix except the background $U$ with replacement of $W$
\begin{eqnarray}\label{18_7_24_1}
D= m + c_N^\prime D_N[W] +  (X_{n,\mu}\delta_{n+\hat\mu,m} + X_{n,-\mu}\delta_{n-\hat\mu,m})
\end{eqnarray}

Similar to (\ref{18_7_18_21_1}), we want to calculate
\begin{eqnarray}\label{18_1_29_20}
 &&  Y^\dagger_{n,a}  \frac{\partial  D_{nm;ab}}{\partial \omega^{i}_{k,\rho}}  X_{m,b}  \nonumber \\
 &=& Y^\dagger_{n,a} \frac{\partial  D_{nm;ab}}{\partial W_{p,\nu;cd}}
  \frac{\partial  W_{p,\nu;cd}}{\partial V_{p,\nu;ef}} \frac{\partial  V_{p,\nu;ef}}{\partial U_{l,\mu;gh}}
   \frac{\partial  U_{l,\mu;gh}}{\partial \omega^{i}_{k,\rho}}   X_{m,b}  \nonumber \\
 &=& F_{p,\nu;ef}\frac{\partial  V_{p,\nu;ef}}{\partial U_{l,\mu;gh}}   \delta_{lk}\delta_{\mu \rho} (\text{i}T_iU_{l,\mu})_{gh}   \nonumber \\
 &=& F_{p,\nu;ef}\frac{\partial  V_{p,\nu;ef}}{\partial U_{k,\rho;gh}}  (\text{i}T_iU_{k,\rho})_{gh}   \nonumber \\
 &=& \tilde F_{k,\rho;gh}  (\text{i}T_iU_{k,\rho})_{gh}   \nonumber \\
 &=&   \text{tr} (\text{i}T_i B )
\end{eqnarray}
where $a,b,\cdots,h$ denotes the color indices. Here we introduced
\begin{eqnarray}\label{18_7_24_4}
F_{p,\nu;ef} = \Big(Y_{n,a}^\dagger  \frac{\partial  D_{nm;ab}}{\partial W_{p,\nu;cd}}
     X_{m,b} \Big)\frac{\partial  W_{p,\nu;cd}}{\partial V_{p,\nu;ef}}
\end{eqnarray}
\begin{eqnarray}\label{18_7_24_5}
\tilde F_{k,\rho;gh} = F_{p,\nu;ef}\frac{\partial  V_{p,\nu;ef}}{\partial U_{k,\rho;gh}}
\end{eqnarray}
and
\begin{eqnarray}\label{18_7_24_6}
B = U_{k,\rho} \tilde F_{k,\rho}^T
\end{eqnarray}
The calculation in the bracket of (\ref{18_7_24_4}) is the same with the those in asqtad fermion (See section \ref{asqtad}) where the background gauge field $U$ is replaced by $W$. By the definition of $V$ in (\ref{18_6_6_7_2}), $\tilde F$ in (\ref{18_7_24_5}) can be written as
\begin{eqnarray}\label{18_7_24_7}
\tilde F_{k,\rho;gh} &=& F_{p,\nu;ef} \sum_P \tilde c_P  \frac{\partial  }{\partial U_{k,\rho;gh}} ( U_{n_1,d_1}\cdots  U_{n_L,d_L} )_{ef} \nonumber \\
 &=& \sum_P \tilde c_P  \sum_{l=1}^L F_{p,\nu;ef} \Big(U_{n_1,d_1}\cdots U_{n_{l-1},d_{l-1}}\frac{\partial U_{n_{l},d_{l}} }{\partial U_{k,\rho;gh}}  U_{n_{l+1},d_{l+1}} \cdots U_{n_L,d_L} \Big)_{ef} \nonumber \\
 &=& \sum_P \tilde c_P  \sum_{l=1}^L  \text{tr} \Big(U_{n_1,d_1}\cdots U_{n_{l-1},d_{l-1}}\frac{\partial U_{n_{l},d_{l}} }{\partial U_{k,\rho;gh}}  U_{n_{l+1},d_{l+1}} \cdots U_{n_L,d_L} F_{p,\nu}^T \Big)
\end{eqnarray}
where the sum over $p, \nu$ should be understood. The sum over $P$ runs for all paths starting from some site and ends at the neighboring site in the positive direction. Let $P$ be a path of length $L$, connecting $L+1$ sites $(n_i)_{i=1}^L$, with $n_{i+1}=n_i+\hat d_i$, $i=1,\cdots,L-1$. Here $n_1=p$, $n_L+\hat d_L = p+\hat \nu$. Then

\begin{eqnarray*}
\frac{\partial U_{n_{l},d_{l}} }{\partial U_{k,\rho;gh}} =   \left\{ \begin{array}{ll}
\delta_{n_l,k} O_{gh}, &   \text{if } d_l = \rho \\
\delta_{n_l-\hat\rho,k} O_{gh}, & \text{if } d_l = -\rho \\
0, &  \text{otherwise }  \\
\end{array} \right.
\end{eqnarray*}
where $O_{gh}$ is a $3\times 3$ matrix with 1 at $(g,h)$ and 0, otherwise.

The calculation of $\tilde F$ can be written as follows.
\begin{enumerate}
\setlength{\parskip}{-3pt}
\item For all direction $\rho$
\item \hspace{0.5cm} $\tilde F_{k,\rho} =0$ for each site $k$
\item \hspace{0.5cm} For all path $P$
\item \hspace{1.2cm} For each link $l$ of $P$
\item \hspace{1.9cm} If $d_l\neq \pm \rho$, go to step 4
\item \hspace{1.9cm}   For all $(p, \nu)$, calculate $ A_{p,\nu} =U_{n_1,d_1}\cdots U_{n_{l-1},d_{l-1}}$ and
$B_{p,\nu} =  U_{n_{l+1},d_{l+1}} \cdots U_{n_L,d_L} F_{p,\nu}^T$
\item  \hspace{1.9cm}  $\tilde F_{k,\rho;gh} += \sum_{p,\nu} \sum_a A_{p,\nu;ag}B_{p,\nu;ha}$, where $k=n_l$ if $d_l=\rho$;  $k=n_l-\hat\rho$ if $d_l=-\rho$
\end{enumerate}

\section{Rooted staggered fermion}\label{rooted}
\begin{figure}[h]
\centering
\includegraphics[width=9cm,height=7cm]{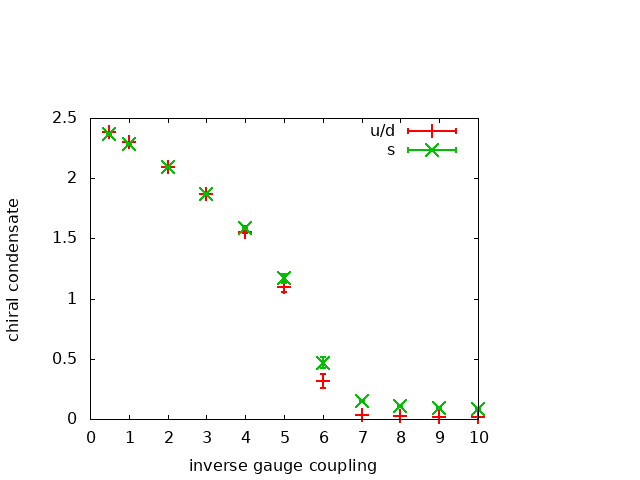}
\caption{The dependence of chiral condensate on inverse gauge coupling $\beta$.} \label{21_12_5_1}
\end{figure}

Now we consider the rational HMC for rooted staggered fermion action, which caused rather controversial discussion. See the comments of the rooted staggered fermion \cite{Bazavov_1349}. After integrating out the Grassman-valued quark
fields, the 2+1 quark flavor QCD partition function is
given by a functional integral over gauge fields,
\begin{eqnarray}\label{18_3_14_1}
Z= \int DU \det(M_l)^{\frac{2}{4}}\det(M_s)^{\frac{1}{4}} e^{-S_G} = \int DU \det(M_l^{\frac{2}{4}}M_s^{-\frac{2}{4}})\det(M_s^{\frac{3}{4}}) e^{-S_G} =
\int DU D\phi_l D\phi_s  e^{-S_G-S_F}
\end{eqnarray}
where the two degenerated up/down (light) fermion $l$ with bare mass $m_l$ and a stranger fermion $s$ with bare mass $m_s$ are considered. The staggered fermion matrix for fermion $l$ ($s$)
\begin{eqnarray}\label{19_6_21_1}
M_{l/s}=D_{eo}D_{eo}^\dagger+4m_{l/s}^2 \equiv M_0+4m_{l/s}^2
\end{eqnarray}
 is the fermion matrix, where $D_{eo}$ is given by
 (\ref{18_7_19_3_1}).
The staggered fermion matrix $M_l$ ($M_s$) and the pseudofermion $\phi_l$ ($\phi_s$) are both defined on the even sites.
The determinants in (\ref{18_3_14_1}) represent the (l)ight and (s)trange
quark contributions to the vacuum. The fermion action in (\ref{18_3_14_1}) is 
\begin{eqnarray}\label{18_3_14_2}
S_F = \phi_l^\dagger (M_l^{-\frac{2}{4}}M_s^{\frac{2}{4}}) \phi_l +\phi_s^\dagger M_s^{-\frac{3}{4}} \phi_s
\end{eqnarray}
In the hybrid Monte Carlo method, $\phi_l$ and $\phi_s$ are sampled according to the distribution $e^{-S_F}$, which can be realized according to \begin{eqnarray}\label{18_3_14_2_1}
\phi_l = M_l^{\frac{2}{8}}M_s^{-\frac{2}{8}} \eta, \quad \phi_s = M_s^{\frac{3}{8}}\eta
\end{eqnarray}
where $\eta$ is sampled from the Gaussian distribution $e^{-\eta^\dagger \eta}$. $M_l^{\frac{2}{8}}M_s^{-\frac{2}{8}}$  can be approximated by
by rational polynomial of $M_0$
$$ M_l^{\frac{2}{8}}M_s^{-\frac{2}{8}}  = (M_0+4m_l)^{\frac{2}{8}}(M_0+4m_s)^{-\frac{2}{8}}  \approx \tilde\alpha_0 + \sum_{p=1}^N  \tilde\alpha_p(M_0+\tilde\beta_p)^{-1} $$
Similarly, $M_s^{\frac{3}{8}}$ can also be approximated by another rational polynomial of $M_0$. 
The fermion action in (\ref{18_3_14_2}) can be calculated by
\begin{eqnarray}\label{18_3_14_3}
S_F = ( M_l^{-\frac{2}{8}}M_s^{\frac{2}{8}} \phi_l)^\dagger ( M_l^{-\frac{2}{8}}M_s^{\frac{2}{8}} \phi_l) +(M_s^{-\frac{3}{8}} \phi_s)^\dagger (M_s^{-\frac{3}{8}} \phi_s)
\end{eqnarray}
where $M_l^{-\frac{2}{8}}M_s^{\frac{2}{8}}$ and $M_s^{-\frac{3}{8}}$ are also be approximated by rational polynomials of $M_0$.  
In order to calculate the fermion force, each term (e.g., $\phi_l^\dagger (M_l^{-\frac{2}{4}}M_s^{\frac{2}{4}}) \phi_l $) of the fermion action in (\ref{18_3_14_2}) is approximated as 
\begin{eqnarray}\label{18_3_14_5}
  \phi_l^\dagger (M_l^{-\frac{2}{4}}M_s^{\frac{2}{4}}) \phi_l\approx
\phi_l^\dagger \Big( \alpha_0 +  \sum_{p=1}^M \alpha_p (M_0 + \beta_p)^{-1} \Big)\phi_l
\end{eqnarray}
with its derivative
\begin{eqnarray}\label{18_3_14_6}
\frac{\partial}{\partial \omega_{k,\rho}^i}\Big(\phi_l^\dagger (M_l^{-\frac{2}{4}}M_s^{\frac{2}{4}}) \phi_l\Big) & \approx & -\sum_{p=1}^M \alpha_p  x_p ^\dagger \frac{\partial  M_0}{\partial \omega_{k,\rho}^i} x_p \nonumber \\
 & = & -\sum_{p=1}^M \alpha_p  x_p ^\dagger \Big[  \frac{\partial D_{eo}}{\partial \omega_{k,\rho}^i}  D_{eo}^\dagger + \text{c.c.}\Big] x_p \nonumber \\
 & = & -\sum_{p=1}^M \alpha_p  x_p ^\dagger   \frac{\partial D_{eo}}{\partial \omega_{k,\rho}^i}  x_p^o + \text{c.c.} \nonumber \\
 & = & -\sum_{p=1}^M \alpha_p  x_{p,n}^\dagger   \frac{\partial D_{eo,n,m}}{\partial \omega_{k,\rho}^i}  x_{p,m}^o + \text{c.c.} \nonumber \\
 & = &- \sum_{\text{all 688 loops } P \text{ in asqtad}} c_P \sum_{p=1}^M \alpha_p  x_{p,n}^\dagger \Big[ U_1\cdots U_{l-1}\frac{\partial  U_l}{\partial \omega^{i}_{k,\rho}} U_{l+1}\cdots U_L \Big]  x_{p,m}^o + \text{c.c.}
\end{eqnarray}
where the path $P= U_1\cdots U_L$ is a length of $L$ from lattice site $n$ to $m$, $x_p$ and  $x_p^o$ are given by 
$$ (M_0+\beta_p) x_p = \phi_l, \quad  x_p^o = D_{eo}^\dagger x_p , \quad p =1,\cdots,M  $$
Here $x_p$ and $x_p^o$ are defined on the even sites and odd sites, respectively. 
If $U_l = U_{k,\rho} $ in (\ref{18_3_14_6}), 
\begin{eqnarray}\label{18_3_14_6_1}
\frac{\partial}{\partial \omega_{k,\rho}^i}\Big(\phi_l^\dagger (M_l^{-\frac{2}{4}}M_s^{\frac{2}{4}}) \phi_l\Big)
=- \sum_{\text{all 688 loops } P \text{ in asqtad}} c_P \text{tr} \Big( \text{i}T_iU_l\cdots U_L \Big[ \sum_{p=1}^M \alpha_p  (x_{p,m}^o \otimes x_{p,n}^\dagger)\Big]
U_1\cdots U_{l-1}\Big) + \text{c.c.}
\end{eqnarray}
If $U_l = U_{k,\rho}^\dagger $ in (\ref{18_3_14_6}),
\begin{eqnarray}\label{18_3_14_6_2}
\frac{\partial}{\partial \omega_{k,\rho}^i}\Big(\phi_l^\dagger (M_l^{-\frac{2}{4}}M_s^{\frac{2}{4}}) \phi_l\Big)
=- \sum_{\text{all 688 loops } P \text{ in asqtad}} c_P \text{tr} \Big( \text{i}T_iU_l^\dagger \cdots U_1^\dagger \Big[ \sum_{p=1}^M \alpha_p  (x_{p,m}^o \otimes x_{p,n}^\dagger)^\dagger\Big]
U_L^\dagger\cdots U_{l+1}^\dagger\Big) + \text{c.c.}
\end{eqnarray}
We use the rational HMC algorithm to simulate the chiral condensate of up/light fermion and strange fermion, where $m_l=0.01$, $m_s=0.05$,
with the lattice size 8 in each direction, $u_0=0.862$.
FIG. \ref{21_12_5_1} shows the chiral condensate of the light(u/d) fermion and strange fermion on the inverse gauge coupling $\beta$. 
The chiral condensate decreases with increasing $\beta$.  

\section{Smeared fermion}\label{smeared}

Both the asqtad and HISQ fermion introduce smeared gauge field to reduce the taste violation of the standard staggered fermion. In fact the other fermions can also benefit from smearings of gauge field. For Wilson fermions the spread of the near zero real modes of the Wilson Dirac operator make it impossible to simulate at small quark masses without going to very fine lattice spacing or large volumes. Smearing can reduce the spread of the eigenvalues \cite{DeGrand_259}. Chiral fermions including overlap fermion and domain wall fermion also benefit from smeared gauge field. This is because the smearing reduces the occurrence of low modes of the Kernel operator from which it is constructed
\cite{DeGrand_034507}\cite{DeGrand_054503}. But one cannot perform this smearing too aggressively, however, since the smearing may distort short distance physics and enhance discretization errors. Here I give the Hyper-cubic (HYP) blocking smearing \cite{Hasenfratz_034504}\cite{Hasenfratz_029}.

HYP smearing consist three steps of projected APE type smearing.
\begin{eqnarray}\label{18_7_28_0}
V_{n,\mu}& = & \text{Proj}  \Big[(1-\alpha_1) U_{n,\mu} + \frac{\alpha_1}{6} \sum_{\pm \nu \neq \mu} \tilde V_{n,\nu;\mu}\tilde V_{n+\hat\nu,\mu;\nu}\tilde V^\dagger_{n+\hat\mu,\nu;\mu}\Big] \nonumber \\
\tilde V_{n,\mu;\nu}& = & \text{Proj}  \Big[(1-\alpha_2) U_{n,\mu} + \frac{\alpha_2}{4} \sum_{\pm \rho \neq \nu, \mu} \bar V_{n,\rho;\nu\mu}\bar V_{n+\hat\rho,\mu;\rho\nu}\bar V^\dagger_{n+\hat\mu,\rho;\nu\mu}\Big] \nonumber \\
\bar V_{n,\mu;\nu\rho}& = & \text{Proj}  \Big[(1-\alpha_3) U_{n,\mu} + \frac{\alpha_3}{2} \sum_{\pm \eta \neq \rho,\nu,\mu} U_{n,\eta}U_{n+\hat\eta,\mu}U^\dagger_{n+\hat\mu,\eta}\Big]
\end{eqnarray}
where we used the notations in Ref. \cite{Hasenfratz_029}. Here $\text{Proj}  $ denotes the projection to $\text{SU(3)}$ matrix.
$V_{n,\mu}$ is the smeared link from the site $n$ in direction $\mu$ while $U_{n,\mu}$ is the original (thin) link.  From the definition of $\bar V$ the two indices $\nu,\rho$ in $\bar V_{n,\mu;\nu\rho}$ can be interchanged, and  $\bar V_{n,\mu;\nu\rho}$ can be defined for $\nu<\rho$ and $\nu,\rho\neq \mu$, and thus there are 12 combinations for $(\mu;\nu\rho)$. In practical implementation of HYP, we introduce
$W_{n,\mu,\eta} = \bar V_{n,\mu;\nu\rho}$ for $\eta  = \overline{\mu\nu\rho}  \neq \mu,\nu,\rho$. The second and third step of HYP smearing can be rewritten as
\begin{eqnarray}\label{18_7_28_1}
\tilde V_{n,\mu;\nu}& = & \text{Proj}  \Big[(1-\alpha_2) U_{n,\mu} + \frac{\alpha_2}{4} \sum_{\pm \rho \neq \nu, \mu, \lambda = \overline{\rho\nu\mu}} W_{n,\rho,\lambda}W_{n+\hat\rho,\mu,\lambda}W^\dagger_{n+\hat\mu,\lambda}\Big] \nonumber \\
W_{n,\mu,\eta}& = & \text{Proj}  \Big[(1-\alpha_3) U_{n,\mu} + \frac{\alpha_3}{2} \Big( U_{n,\eta}U_{n+\hat\eta,\mu}U^\dagger_{n+\hat\mu,\eta} + (\eta\rightarrow -\eta)\Big)\Big]
\end{eqnarray}
For each site $n$, we want to store $\tilde V_{n,\mu;\nu}$ for $\mu\neq \nu$ and $W_{n,\mu,\eta}$ for $\mu\neq \eta$.

The smeared Wilson fermion action can be written as $S_F = \phi^\dagger (D^\dagger D)^{-1}\phi  $ where $\phi$ is the pseudo-fermion and $D$
is the smeared Wilson matrix, i.e., the thin link $U_{n,\mu}$ is replaced by the smeared link $V_{n,\mu}$.
By the chain rule, one has
\begin{eqnarray}\label{18_7_28_1_1}
 \frac{\partial S_F}{\partial \omega}  = \text{Retr} \Big(\Sigma_{n,\mu} \frac{\partial   V_{n,\mu}}{\partial \omega} \Big)
\end{eqnarray} 
 \begin{eqnarray}\label{18_7_28_2}
 \Sigma_{n,\mu} \frac{\partial   V_{n,\mu}}{\partial \omega}  & = & \Sigma_{n,\mu} \Big[ \frac{\partial V_{n,\mu}}{\partial U_{n,\mu}}  \frac{\partial  U_{n,\mu}}{\partial \omega} +
\frac{\partial V_{n,\mu}}{\partial \tilde V_{m,\nu;\rho}}  \frac{\partial   \tilde V_{m,\nu;\rho} }{\partial \omega} \Big]  \nonumber \\
&=& \Sigma^{(1)}_{\mu}  \frac{\partial  U_{\mu} }{\partial \omega} + \tilde \Sigma^{(1)}_{m,\nu;\rho} \Big[  \frac{\partial  \tilde V_{m,\nu;\rho}}{\partial U_{m,\nu}}  \frac{\partial  U_{m,\nu} }{\partial \omega} +
 \frac{\partial  \tilde V_{m,\nu;\rho}}{\partial \bar V_{n,\alpha;\beta\gamma}} \frac{\partial \bar V_{n,\alpha;\beta\gamma}  }{\partial \omega}  \Big]  \nonumber \\
&=& \Sigma^{(1)}_{\mu}  \frac{\partial  U_{\mu} }{\partial \omega} + \Sigma^{(2)}_{\nu}  \frac{\partial  U_{\nu} }{\partial \omega} +\tilde \Sigma^{(2)}_{n,\alpha;\beta\gamma}   \frac{\partial\bar V_{n,\alpha;\beta\gamma}}{\partial U_{m,\nu}}  \frac{\partial  U_{m,\nu} }{\partial \omega}  \nonumber \\
&=&  \Big(\Sigma^{(1)}_{\nu}  + \Sigma^{(2)}_{\nu}  +  \Sigma^{(3)}_{\nu} \Big)  \frac{\partial  U_{\nu} }{\partial \omega}
\end{eqnarray}
where
\begin{eqnarray}\label{18_7_28_3}
\Sigma_{n,\mu}  = \frac{\partial   S_F}{\partial V_{n,\mu}}, \quad \Sigma^{(1)}_{n,\mu}  = \Sigma_{n,\mu}\frac{\partial V_{n,\mu}}{\partial U_{n,\mu}}, \quad \tilde \Sigma^{(1)}_{m,\nu;\rho} = \Sigma_{n,\mu}\frac{\partial V_{n,\mu}}{\partial \tilde V_{m,\nu;\rho}} , \quad
\end{eqnarray}
\begin{eqnarray}\label{18_7_28_4}
\Sigma^{(2)}_{m,\nu} = \tilde \Sigma^{(1)}_{m,\nu;\rho} \frac{\partial  \tilde V_{m,\nu;\rho}}{\partial U_{m,\nu}}, \quad
\tilde \Sigma^{(2)}_{n,\alpha;\beta\gamma} = \tilde \Sigma^{(1)}_{m,\nu;\rho}\frac{\partial  \tilde V_{m,\nu;\rho}}{\partial \bar V_{n,\alpha;\beta\gamma}}, \quad \Sigma^{(3)}_{m,\nu} = \tilde \Sigma^{(2)}_{n,\alpha;\beta\gamma}   \frac{\partial\bar V_{n,\alpha;\beta\gamma}}{\partial U_{m,\nu}}
\end{eqnarray}
Inserting (\ref{18_7_28_2}) to (\ref{18_7_28_1_1}), one has
$$ \frac{\partial S_F}{\partial \omega^i_{k,\rho}}  = \text{Retr}  \Big(  (iT_i)
U_{k,\rho} (  \Sigma^{(1)}_{k,\rho}  + \Sigma^{(2)}_{k,\rho}  +  \Sigma^{(3)}_{k,\rho}) \Big)$$
The details of calculations for $\Sigma^{(i)}$ can be found in \cite{Hasenfratz_029}.

\section{Staggered Wilson fermion}\label{staggered_wilson}
Adams introduced the massless Staggered Wilson fermion matrix \cite{Adams_141602}\cite{Adams_394}
\begin{eqnarray}\label{17_6_29_1}
D_{\text{sw}} = D_{\text{st}} +  W_{\text{st}}
\end{eqnarray}
where $D_{\text{st}} = \frac{1}{2} D_1$ is the massless staggered fermion matrix and $D_1$ is given by (\ref{18_6_6_3_1}).
\begin{eqnarray}\label{17_6_29_2}
W_{\text{st}} = r(1- \varepsilon\Gamma_5)
\end{eqnarray}
with Wilson-like parameter $r>0$, $\varepsilon \psi_n = (-1)^{(n_1+n_2+n_3+n_4)}\psi_n$.
$\Gamma_5 = \eta_5 C$ where $\eta_5 = \eta_1\eta_2\eta_3\eta_4$, i.e., $\eta_5 \psi_n = (-1)^{(n_1+n_3)}\psi_n
$. The operator $C$ is given by
\begin{eqnarray}\label{17_6_29_4}
C =  \frac{1}{4!}\sum_{\mu\nu\lambda\sigma}C_\mu C_\nu C_\lambda C_\sigma
\end{eqnarray}
where the sum $\sum_{\alpha\beta\gamma\delta}$ runs for all $4!$ permutations of $(1,2,3,4)$ and
\begin{eqnarray}\label{17_6_29_5}
C_\mu = \frac{T_{+\mu} + T_{-\mu} }{2}
\end{eqnarray}
with $T_{\pm\mu}  \psi_n = U_{n,\pm\mu}\psi_{n\pm\hat\mu}$. Obviously,
\begin{eqnarray}\label{17_6_29_6}
(\varepsilon D_{\text{sw}})^\dagger =\varepsilon D_{\text{sw}} \Longleftrightarrow
  D_{\text{sw}}^\dagger =\varepsilon D_{\text{sw}} \varepsilon
\end{eqnarray}
 due to $\varepsilon C_\mu = C_\mu \varepsilon$. 

The fermion action for the staggered Wilson fermion with two degenerate flavors is
\begin{eqnarray}\label{18_7_26_14}
S_F = \phi^\dagger (D_{\text{sw}}^\dagger D_{\text{sw}})^{-1} \phi
\end{eqnarray}
with pseudofermion fields $\phi$. To calculate the fermion force, we have to compute
$$Y^\dagger  \frac{\partial D_{\text{sw}}}{\partial \omega^{i}_{k,\rho}}  X  = Y^\dagger  \frac{\partial D_{\text{st}}}{\partial \omega^{i}_{k,\rho}}  X  +Y^\dagger  \frac{\partial W_{\text{sw}}}{\partial \omega^{i}_{k,\rho}}  X   $$
 with $X =  (D_{\text{sw}}^\dagger D_{\text{sw}})^{-1}\phi$ and  $Y = D_{\text{sw}} X$. The calculation of
 $Y^\dagger  \frac{\partial D_{\text{st}}}{\partial \omega^{i}_{k,\rho}}  X$ is given before. Since the operator $\varepsilon$ and $\eta_5$ are diagonal in lattice space, which does not depend on $U$,
\begin{eqnarray}\label{18_7_26_15}
&& Y^\dagger  \frac{\partial W_{\text{sw}}}{\partial \omega^{i}_{k,\rho}}  X  \nonumber \\
& = & - \frac{r}{4!} \sum_{\mu\nu\rho\sigma}  ( Y^\dagger \varepsilon \eta_5)  \frac{\partial C_\mu} {\partial \omega^{i}_{k,\rho}} (C_\nu C_\rho C_\sigma X) + \cdots  \nonumber \\
& = & - \frac{r}{2\times 4!} \sum_{\mu\nu\lambda\sigma}  \tilde Y_n^\dagger \Big[  \frac{\partial U_{n,\mu}} {\partial \omega^{i}_{k,\rho}} \tilde X_{n+\hat\mu} + \frac{\partial U_{n-\hat\mu,\mu}^\dagger} {\partial \omega^{i}_{k,\rho}} \tilde X_{n-\hat\mu}\Big]+ \cdots  \nonumber \\
& = & - \frac{r}{2\times 4!} \sum_{\mu\nu\lambda\sigma}  \tilde Y_n^\dagger \Big[ \delta_{n,k}\delta_{\mu,\rho} \text{i} T_i U_{k,\rho} \tilde X_{n+\hat\mu} + \delta_{n-\hat\mu,k}\delta_{\mu,\rho}  U_{k,\rho}^\dagger(-\text{i} T_i) \tilde X_{n-\hat\mu}\Big]+ \cdots  \nonumber \\
& = & - \frac{r}{2\times 4!} \sum_{\nu\lambda\sigma} \Big[  \tilde Y_k^\dagger    \text{i} T_i U_{k,\rho} \tilde X_{k+\hat\rho} +  \tilde Y_{k+\hat\rho}^\dagger   U_{k,\rho}^\dagger(-\text{i} T_i) \tilde X_{k}\Big]+ \cdots  \nonumber \\
& = & - \frac{r}{2\times 4!} \sum_{\nu\lambda\sigma} \text{tr} \Big[  \text{i} T_i \Big(  ( U_{k,\rho} \tilde X_{k+\hat\rho})\otimes \tilde Y_k^\dagger  -  \tilde X_{k}\otimes ( \tilde Y_{k+\hat\rho}^\dagger   U_{k,\rho}^\dagger)\Big)\Big]+ \cdots
\end{eqnarray}
where $\tilde Y^\dagger =Y^\dagger \varepsilon \eta_5$ and $\tilde X =C_\nu C_\rho C_\sigma X$. The other three terms ($\cdots$) in (\ref{18_7_26_15}) can also be written as the first term.

\section{Overlap fermion}\label{overlap}
The overlap fermion matrix is
\begin{eqnarray}\label{18_7_20_1}
D_{\text{ov}} = (1-m)D^0_{\text{ov}} + m
\end{eqnarray}
where $m$ is the non-dimensional fermion mass and
\begin{eqnarray}\label{18_7_20_2}
D^0_{\text{ov}} = \frac{1}{2}\Big(1+\gamma_5 \text{sign}[H]\Big)
\end{eqnarray}
 is the overlap fermion matrix at $m=0$,
satisfying the Ginsparg-Wilson equation \cite{Neuberger_177}. $\text{sign}[H] \equiv H(H^\dagger H)^{-1/2}$ where the Hermitian matrix $H$ is
\begin{eqnarray}\label{18_7_20_2_1}
H = \gamma_5 (D_{\text{w}}^0 -  M)
\end{eqnarray}
Here $M>0$ is the large mass and $D_{\text{w}}^0$ is the Wilson fermion matrix  at chiral limit $m=0$
\begin{eqnarray}\label{18_7_20_3}
D_{\text{w};n,m}^0 = \frac{1}{2}\sum_{\mu= 1}^{4}\Big( (1-\gamma_\mu)  U_{n,\mu}\delta_{n+\hat\mu,m}+
 (1+\gamma_\mu)  U_{n,-\mu}\delta_{n-\hat\mu,m}\Big)
\end{eqnarray}
The numerical implementation requires an approximation of the matrix sign function of a Wilson-like fermion operator, and
various approaches are being used. In fact, it is possible to rewrite these approximations in terms
of a five-dimensional formulation, showing that the domain wall fermion and overlap fermion are essentially
equivalent \cite{Borici_9912040}\cite{Kennedy_0607038}.

By introducing the pseudo-fermion $\phi$, the fermion action is
\begin{eqnarray}\label{18_7_20_4}
S_F = \phi^\dagger (D_{\text{ov}}^\dagger D_{\text{ov}})^{-1}\phi
\end{eqnarray}
The derivative of the fermion action is the same with (\ref{18_6_24_0}) where the fermion matrix is replaced by the overlap fermion matrix $D_{\text{ov}}$. To calculate the $X$ and $Y$ in (\ref{18_6_24_0_1}), we approximate $\text{sign}[H]$ in (\ref{18_7_20_2}) (See P. 180 in \cite{Gattringer_2009})
\begin{eqnarray}\label{18_7_20_5}
\text{sign}[H] \approx H \sum_{n=0}^{N-1} c_n T_n(\tilde H)
\end{eqnarray}
with
\begin{eqnarray}\label{18_7_20_6}
\tilde H = \frac{2H^2 - (\beta^2 + \alpha^2)}{\beta^2 - \alpha^2}
\end{eqnarray}
where $\alpha$ and $\beta$  are the (in magnitude) smallest and the largest eigenvalues of $H$, respectively. $T_n$ is the first kind of Chebyshev polynomials of
order $n$ and the coefficient $c_n$ in (\ref{18_7_20_5}) is
$$ c_n = \int_{-1}^1 {\rm d}x \frac{r(x)T_n(x)}{\sqrt{1-x^2}}, \quad r(x) = \Big(\frac{1}{2}(\beta^2 + \alpha^2) +
\frac{x}{2}(\beta^2  - \alpha^2)\Big)^{-1/2}  $$
Expanding the RHS of (\ref{18_7_20_5}), one has
\begin{eqnarray}\label{18_7_20_7}
\text{sign}[H] \approx d_1H + d_3H^3 + \cdots + d_{2N-1}H^{2N-1}
\end{eqnarray}
where $\{d_{2l-1}\}_{i=1}^N$ etc., depend on the coefficients $\{c_n\}_{n=0}^{N-1}$, $\alpha$ and $\beta$.

Similar to (\ref{18_7_18_21_1}), we want to calculate
\begin{eqnarray}\label{18_7_20_8}
 &&  Y_n^\dagger  \frac{\partial  D_{\text{ov};n,m}}{\partial \omega^{i}_{k,\rho}}  X_m  \nonumber \\
 &=&\frac{1-m}{2} \sum_{l=1}^N d_{2l-1} \sum_{j=1}^{2l-1}  Y_n^\dagger \gamma_5  H^{j-1} \frac{\partial H_{n,m}}{\partial \omega^{i}_{k,\rho}} H^{2l-1-j} X_m
 \nonumber \\
 &=&\frac{1-m}{2} \sum_{l=1}^N d_{2l-1} \sum_{j=1}^{2l-1}  Y_n^\dagger \gamma_5  H^{j-1} \gamma_5  \frac{1}{2} \Big[(1-\gamma_\rho) (\text{i} T_i) U_{k,\rho}\delta_{n+\hat\rho,m}\delta_{n,k}+
 (1+\gamma_\rho)  U_{k,\rho}^\dagger (-\text{i}T_i) \delta_{n-\hat\rho,m}\delta_{m,k} \Big] H^{2l-1-j} X_m
 \nonumber \\
 &=&\frac{1-m}{4} \sum_{l=1}^N d_{2l-1} \sum_{j=1}^{2l-1}  Y_k^\dagger \gamma_5  H^{j-1} \gamma_5  (1-\gamma_\rho) (\text{i} T_i) U_{k,\rho} H^{2l-1-j} X_{k+\hat\rho} -
 \nonumber \\
 & &\frac{1-m}{4} \sum_{l=1}^N d_{2l-1} \sum_{j=1}^{2l-1}  Y_{k+\hat\rho}^\dagger \gamma_5  H^{j-1} \gamma_5  (1+\gamma_\rho) U_{k,\rho}^\dagger  (\text{i} T_i)  H^{2l-1-j} X_{k}
 \nonumber \\
 &=&   \text{tr} (\text{i}T_i (B - C))+ \text{c.c.}
\end{eqnarray}
where
\begin{eqnarray*}
 B = \frac{1-m}{4} \sum_{l=1}^N d_{2l-1} \sum_{j=1}^{2l-1}\Big(U_{k,\rho} H^{2l-1-j} X_{k+\hat\rho} \Big) \otimes \Big(Y_k^\dagger \gamma_5  H^{j-1} \gamma_5  (1-\gamma_\rho) \Big)
\end{eqnarray*}
\begin{eqnarray*}
C = \frac{1-m}{4} \sum_{l=1}^N d_{2l-1} \sum_{j=1}^{2l-1}\Big( H^{2l-1-j} X_{k} \Big) \otimes \Big( Y_{k+\hat\rho}^\dagger \gamma_5  H^{j-1} \gamma_5  (1+\gamma_\rho) U_{k,\rho}^\dagger\Big)
\end{eqnarray*}
See (\ref{18_1_29_20_2}) and (\ref{18_1_29_20_3}) for comparison of $B$ and $C$.

\section{Domain wall fermion}\label{domain_wall}
The domain wall fermion makes use of a 5D lattice and then construct the chiral Dirac fermions when the lattice size $N_5$ in 5th dimension is large
\cite{{Kaplan_342}}\cite{Shamir_90}\cite{Shamir_2691}\cite{Furman_54}. The domain wall operator can be constructed from the massless Wilson operator $D^0_{\text{w}}$
\begin{eqnarray}\label{17_6_27_12}
&& D_{\text{dw}, ns,mt}  =  \delta_{s,t} ( D^0_{\text{w};n,m} - M_5\delta_{n,m})  + \delta_{n,m} D_{5;s,t}^{\text{dw}}
\end{eqnarray}
with
\begin{eqnarray}\label{17_6_27_13}
D_{5;s,t}^{\text{dw}}  =  \delta_{s,t} - ( 1-\delta_{s,N_5-1})P_-\delta_{s+1,t} -( 1-\delta_{s,0})P_+\delta_{s-1,t} +m (P_-  \delta_{s,N_5-1}\delta_{0,t} +P_+\delta_{s,0}\delta_{N_5-1,t})
\end{eqnarray}
and the chiral projector
\begin{eqnarray}\label{17_6_27_14}
P_\pm = \frac{1\pm \gamma_5}{2}
\end{eqnarray}
Here $M_5$ is the domain wall barrier, $m$ is the bare fermion mass, $s, t = 0,\cdots,N_5-1$ are the indices in the 5th dimension.
The fermion action for the domain wall fermion with two degenerate flavors is
\begin{eqnarray}\label{18_7_20_14}
S_F = \Psi^\dagger \Big(D_{\text{dw}}(m)^\dagger D_{\text{dw}}(m)\Big)^{-1} \Psi  +  \Phi^\dagger \Big(D_{\text{dw}}(1)^\dagger D_{\text{dw}}(1)\Big) \Phi
\end{eqnarray}
with pseudofermion field $\Psi$ or Pauli-Villars field $\Phi$. Here $D_{\text{dw}}(1)$ is the domain wall matrix
$D_{\text{dw}}(m)$ with $m=1$. Since the link variable only appear in the massless Wilson matrix $D^0_{\text{w}}$, the fermion force calculation is
rather simple (See section \ref{Wilson}).

\section{Conclusions}\label{conclusion}
This lecture give the details of the force calculation in lattice QCD. The most popular gauge action and fermion actions, including the one-loop Symanzik improved action, Wilson fermion with clover term, staggered fermion, asqtad fermion, HISQ fermion, smeared fermion, overlap fermion and domain wall fermion. The even-odd precondition in the Wilson fermion and staggered fermions,  are also considered.
The formula for the force can be understand as follows: the derivative of the action can be obtained by inserting $(iT_i)$ or $(-iT_i)$ in the action before $U_{n,\mu}$ or after $U_{n,\mu}^\dagger$ and then combining two parts which are separated by $(iT_i)$ or $(-iT_i)$ in the action.

There are other actions for LQCD, including perfect action \cite{Hasenfratz_9803027}\cite{Bietenholz_921}, D234c action, NRQCD action and Fermilab action.  Alford, Klassen, and Lepage have proposed
improved gauge and Dirac actions, D234c action \cite{Alford_54}, which involves second, third, and
fourth order derivatives. Lepage and collaborators suggested NRQCD action to approximate heavy quark in heavy-heavy
or heavy-light systems \cite{Lepage_196}\cite{Ali_132}. Fermilab action \cite{El-Khadra_3933}, which interpolate between the light improved SW-clover action and the NRQCD for the heavy quark. The force calculations of these action will be considered in the future.  

\clearpage
\vspace{1cm}

 Acknowledgments.
  Daming Li was supported by the
National Science Foundation of China (No. 11271258, 11971309).

{}

\end{CJK*}


\begin{thebibliography}{}

\bibitem{Wilson_2445} K. G. Wilson, Phys. Rev. D 10 (1974), 2445
\bibitem{Creutz_2308}  M. Creutz, Phys. Rev. D 21 (1980), 2308
\bibitem{Montvay} I. Montvay, G. Munster, Quantum Fields on a Lattice, Cambridge University Press, Cambridge, 1994
\bibitem{Gupta_9807028} R. Gupta, Introduction to Lattice QCD, hep-lat/9807028
\bibitem{Rothe} H. J. Rothe, World Sci. Lect. Notes Phys. 74, 1 (2005)
\bibitem{Smit} J. Smit, Cambridge Lect. Notes Phys. 15, 1 (2002)
\bibitem{Gattringer_2009} C. Gattringer and C. Lang, Quantum Chromodynamics on the Lattice: An Introductory Presentation, Lecture Notes in Physics, Springer Berlin Heidelberg, 2009.

\bibitem{Alford_87} M. Alford, W. Dimm, G. P. Lepage, G. Hockney, P. B. Mackenzie,  Phys. Lett. B 361 (1995), 87

\bibitem{Sheikholeslami_216} B. Sheikholeslami and R. Wohlert, Nucl. Phys. B 259, 572 (1985) 216

\bibitem{Luscher_219} M. L\"uscher and P. Weisz, Nucl. Phys. B 479, 429 (1996) 219

\bibitem{Lepage_219} G. P. Lepage and P. B. Mackenzie, Phys. Rev. D 48, 2250 (1993) 219
\bibitem{DeGrand_219} T. DeGrand and C. DeTar, Lattice Methods for Quantum Chromodynamics (World Scientific, Singapore 2006) 219
\bibitem{Jansen_221} K. Jansen and R. Sommer, Nucl. Phys. B 530, 185 (1998) 221

\bibitem{Lepage_074502} G. P. Lepage, Phys. Rev. D 59 (1999), 074502.
\bibitem{Bazavov_1349} A. Bazavov, D. Toussaint, C. Bernard, etc., Rev. Mod. Phys. 82 (2010), 1349

\bibitem{Naik_238} S. Naik, Nucl. Phys. B 316 (1989), 238

\bibitem{Follana_054502} E. Follana, Q. Mason, C. Davies, K. Hornbostel, G. P. Lepage, J. Shigemitsu, H. Trottier, and K. Wong, Phys. Rev. D 75
    (2007), 054502

\bibitem{Davies_114504} C. T. H. Davies et al., Phys. Rev. D 82 (2010), 114504
\bibitem{McNeile_031503} C. McNeile et al., Phys. Rev. D 85 (2012), 031503
\bibitem{Donald_094501} G. C. Donald et al., Phys. Rev. D 86 (2012), 094501 
   
      

\bibitem{DeGrand_259} T. DeGrand, A. Hasenfratz, T. Kovacs, Nucl. Phys. B 547 (1999), 258
\bibitem{DeGrand_034507} T. DeGrand, S. Schaefer, Phys. Rev. D 71 (2005), 034507
\bibitem{DeGrand_054503} T. DeGrand, S. Schaefer, Phys. Rev. D 72 (2005), 054503

\bibitem{Hasenfratz_034504} A. Hasenfratz, F. Knectli, Phys. Rev. D 64 (2005), 034504
\bibitem{Hasenfratz_029} A. Hasenfratz, R. Hoffmann, S. Schaefer, J. High Energy Phys. (2007), 029


\bibitem{Adams_141602} D. H. Adams, Phys. Rev. Lett. 104 (2010) 141602
\bibitem{Adams_394} D. H. Adams, Phys. Lett. B 699 (2011) 394


\bibitem{Neuberger_177} H. Neuberger, Phys. Lett. B 427, 353 (1998) 164
\bibitem{Borici_9912040} A. Borici, NATO Sci. Ser. C 553, 41 (2000), [hep-lat/9912040].
\bibitem{Kennedy_0607038} A. D. Kennedy (2006), [hep-lat/0607038]
\bibitem{Kaplan_342} D. B. Kaplan, Phys. Lett. B 288 (1992) 342
\bibitem{Shamir_90} Y. Shamir, Nucl. Phys. B 406 (1993) 90
\bibitem{Shamir_2691} Y. Shamir: Phys. Rev. Lett. 71 (1993) 2691
\bibitem{Furman_54} V. Furman and Y. Shamir: Nucl. Phys. B 439 (1995) 54

\bibitem{Alford_54} M. Alford, T. Klassen, and P. Lepage, Phys. Rev. D 58 (1998), 034503

\bibitem{Lepage_196} P. Lepage and B. Thacker, Phys. Rev. D 43 (1991) 196; G. Lepage, L. Magnea, C.
Nakhleh, U. Magnea and K. Hornbostel, Phys. Rev. D 46 (1992) 4052

\bibitem{Ali_132} A. Ali Khan, et.al., Phys. Lett. B 427 (1998) 132

\bibitem{Hasenfratz_9803027} P. Hasenfratz, hep-lat/9803027

\bibitem{Bietenholz_921} W. Bietenholz, R. Brower, S. Chandrasekharan, and U.-J. Wiese, Nucl. Phys.
(Proc. Suppl.) B 53 (1997) 921

\bibitem{El-Khadra_3933} A. X. El-Khadra, A. S. Kronfeld, P. B. Mackenzie, Phys. Rev. D 55 (1997) 3933

\end{thebibliography}
\end{document}